\documentstyle[11pt,paspconf,epsf]{article}

\begin{document}

\title{Toward an Understanding of the Baldwin Effect}
\author{Kirk Korista}
\affil{Western Michigan University, Department of Physics, Kalamazoo,
MI 49008}

\begin{abstract}
After a short review of the quasar problem and what we can hope to
learn from their emission line spectra, I discuss the current body of
knowledge concerning quasar emission lines and their relationships to
the local and ionizing continua. I propose a hypothesis that the
Baldwin effect is due to a characteristic relationship between the
continuum spectral energy distribution, the gas metallicity ($Z$), and
the quasar luminosity.  I suggest that such a relationship might arise
naturally from a scenario involving massive galaxy formation and
evolution driving the birth and evolution of the quasar central engines
in terms of four fundamental parameters: $M_{bh}$, $\dot{M}$,
$\dot{M}/M_{bh}$, and $Z$.
\end{abstract}

\keywords{Baldwin Effect, cosmology, galaxy evolution, gas metallicity,
quasar emission lines, quasar evolution, spectral energy distributions}

\section{Introduction}

\subsection{Some history}

Quasars have long been mysterious points of light in our sky.
Originally turning up as star-like radio sources in the early radio
surveys of the late 1950s, their spectra remained a puzzle until
Maarten Schmidt's realization (Schmidt 1963) that their emission lines
are as those seen in emission line galaxies (Seyfert 1943), but greatly
redshifted. The ``discovery object'' was 3C~273 with a redshift of
0.158. This placed quasars billions of light years away and made them
the most distant and luminous entities in the universe. As more were
discovered, the redshift limit was pushed ever higher, meaning that we
were probing ever-earlier epochs in our Universe's history. Soon it was
realized that if we could establish quasars as``standard luminosity
candles'', we could determine their distances independent of their
redshifts and thus determine the expansion history of the Universe.
Thirty-five years later (my lifetime) we are still puzzling over the
story that quasars are telling us.

\subsection{The quasar paradigm}

While a comprehensive understanding of the quasar phenomenon remains
illusive, a standard model has emerged: a supermassive black hole in
the nucleus of a massive galaxy accretes matter at rate $\dot{M}$,
converting gravitational potential energy into light ($L = \eta \dot{M}
c^2$; $\eta \sim 0.1$) and mechanical energy (Blandford \& Rees 1978;
Rees 1984).  Evidence continues to trickle in that quasars represented
a dynamically active stage in the lives of massive galaxies (McLeod \&
Riecke 1995a, 1995b; Bahcall et al.\ 1997; McLure et al.\ 1998), with
the masses of their central black holes related to the host galaxy's
halo mass (Haehnelt, Natarajan, \& Rees 1998; Laor 1998, and these
proceedings).  The rapid appearance of quasars with redshifts $z > 2$
was likely associated with massive galaxy formation (Boyle \& Terlevich
1998; Haehnelt et al.\ 1998; Miller \& Percival 1998; Silk \& Rees
1998). Quasars are now extinct, though their low luminosity cousins,
the Seyfert galaxies, are still found in the here and now.  During
their reign, the quasar ``central engines'' poured out tremendous
numbers of UV/X-ray photons.  In the standard model the mass of the
central black hole and the mass accretion rate together determine the
quasar's luminosity and continuum spectral energy distribution (SED).

\subsection{Quasar emission lines and the stories they tell}

\subsubsection{``Sometimes you can't stick your head in the engine, so
you have to examine the exhaust.'' (Don Osterbrock c.\ 1987).}

The central engine of a quasar is manifested in the accretion of matter
onto a supermassive black hole. This process releases energy, much of
this in the form of light at ionizing energies (e.g., Mathews \&
Ferland 1987) which is generally unobservable due the opacity of the
intergalactic and interstellar media. Therefore, we cannot discern
directly how a large fraction of the energy in the central engine is
dissipated.  However, some of these continuum photons are intercepted
by gas of unknown origin moving rapidly in the black hole's
gravitational potential. These photons ionize and heat this non-LTE
gas, which responds by producing the quasar broad emission line
spectrum.  The recombination lines of Ly$\alpha$ $\lambda$1216 and
He~II $\lambda$1640 are a measure of the numbers of photons lying just
shortward of 912~\AA\/ and 228~\AA\/, respectively.  The ratio of the
intensities of the strongest line coolant, C~IV $\lambda$1549, to the
strongest recombination line, Ly$\alpha$, measures the cooling per
recombination, or the heating per photoionization, and thus reflects
the energy balance of the gas. Since the energy from the continuum
source absorbed by the gas must equal the energy emitted by the gas,
the emission lines serve as an indirect means to measure the ionizing
continuum. This and the likelihood that the emission line widths are
indicative of the central gravitational potential well mean that we can
hope to learn from the emission lines the nature of the central engines
of quasars.

\subsubsection{Origin \& Nature of the Emitting Gas.}

The emission line spectra of quasars must be leaving clues, albeit
apparently subtle ones, as to the origin and nature of the emitting
gas. For a discussion of this see my other contribution to these
conference proceedings.

\subsubsection{Chemical abundances.}

The strengths of the heavy element emission lines in even the highest
redshift quasars give testimony to the heavy element enrichment from
stellar evolution that had to occur within the first billion years of
cosmic history. If they can be understood, quasar emission lines will
elucidate the chemical evolution of galactic nuclei over the time span
of $\sim$~10\% -- 90\% the age of the Universe (Hamann \& Ferland 1993;
Hamann \& Ferland 1999; Hamann, these proceedings).

\subsubsection{Quasar evolution.}

Quasars are known to have evolved in number density and luminosity over
past 10 or so billion years. Their rapid rise at high redshifts
(presumably due to galaxy formation), their reign near $z \approx 2.5$,
and their demise at lower redshifts, as the source of gas ``ran dry''
and/or the accretion mechanism itself failed (Haiman \& Menou 1998), is
reasonably well established (Boyle, Shanks, \& Peterson 1988; Boyle
1991; Hewett, Foltz, \& Chaffee 1993; Goldschmidt \& Miller 1997; La
Franca \& Cristiani 1997; Graham, Clowes, \& Campusano 1998). Quasar
evolution may be tied to the dynamical and chemical evolution of
massive galaxies ($\S~1.2$). But we do not understand these things at
the fundamental level. Quasar emission lines, mirroring the action
occurring in galactic nuclei, are keys to unlocking these mysteries.

\subsubsection{Beacons of cosmic history?}
Since the 1970s we've known of the so-called ``Ly$\alpha$'' forest of
absorption lines in the spectra of quasars representing matter along
the line of sight to the quasar. The statistics in number and character
of the Ly$\alpha$ forest demonstrate an evolution with look-back time
in the gas that may represent galaxies, protogalaxies, or failed
galaxies. However, quasars themselves are strewn throughout most of
cosmic history and their intrinsic light must be telling us something
about that history. Their connection to galaxy evolution is an
example.  Another is if their light could be somehow established as a
standard luminosity candle, we might establish the expansion history
and fate of the Universe. Detailed analysis of quasar spectra became
possible in the late 1970s with the advent of sensitive electronic
spectrographs. It was at this time that a young astronomer by the name
of Jack Baldwin was compiling spectral characteristics of quasars, when
he came across something interesting.

\subsection{Observed line, continuum, \& luminosity correlations}

This is a brief summary of the situation. For a more complete review
the reader is directed to the conference proceedings contribution from
Osmer \& Shields.

\subsubsection{The Baldwin Effect.}

Uncharacteristic of Galactic emission line sources (e.g., novae, PNe,
H~II regions), the spectra of quasars are remarkably homogeneous. The
same emission lines sitting atop a ``blue bump'' continuum, with
similar intensity ratios, are observed again and again --- this, in the
face of 4--5 orders of magnitude in their luminosities and billions of
years of lookback time. However, upon comparison of the C~IV emission
line flux to that of the underlying continuum flux density
$f_{\lambda}$, Jack discovered that this ratio is in inverse proportion
to the monochromatic ultraviolet continuum luminosity $L_{uv}$
(Baldwin 1977). \begin{equation} \log W_{\lambda}(C~IV) \propto
\beta\log L_{uv}, ~\beta < 0 \end{equation} This effect is illustrated
dramatically, though qualitatively, in Figure~1. If calibrated, a
measurement of a quasar's C~IV equivalent width would indicate its
luminosity and thus its luminosity distance.

\begin{figure}
\plotfiddle{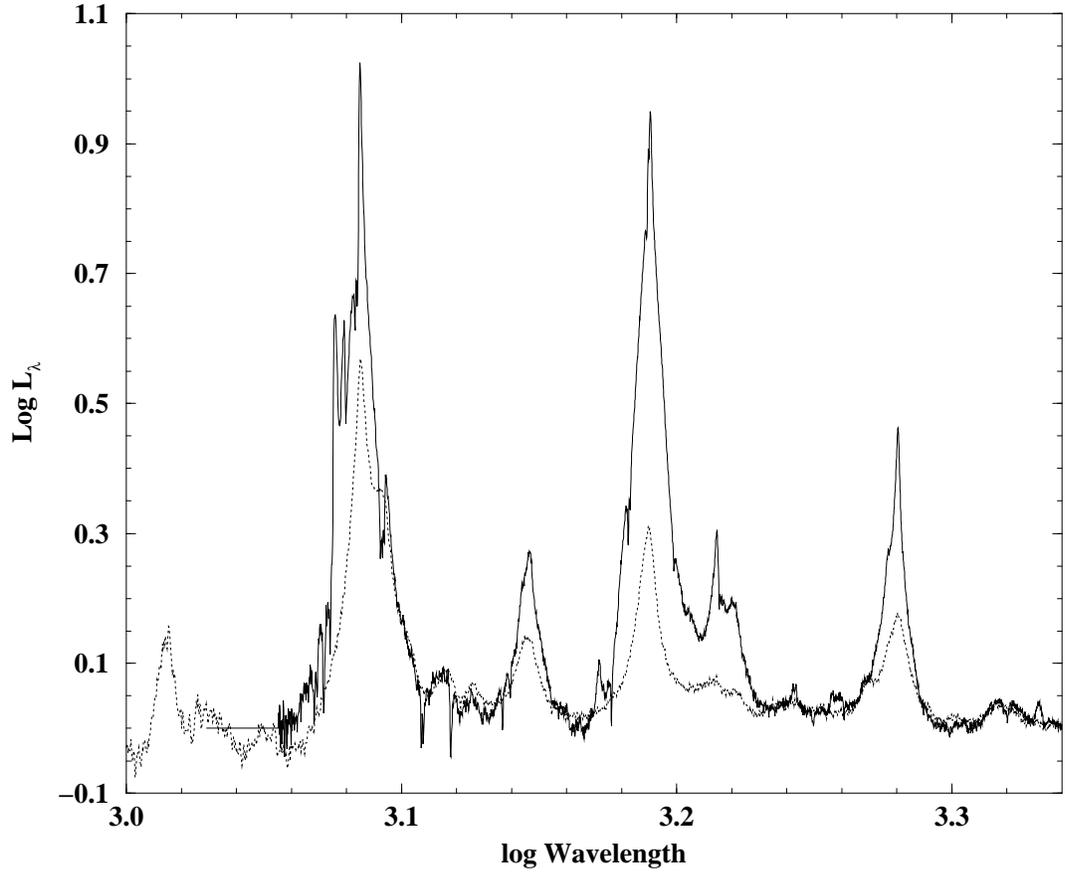}{3.0in}{270.}{70}{70}{-250}{400}
\caption{A qualitative illustration of the Baldwin effect. The solid
curve is the continuum normalized mean UV spectrum of the Seyfert~1
nucleus of NGC~5548 ($z = 0.0174$) from the 1993 monitoring campaign
(Korista et al.\ 1995). The dotted curve is a portion of the revised
version of the composite quasar spectrum from the Large Bright Quasar
Survey (Morris 1994), also continuum normalized. Effectively $\sim$~3
orders of magnitude in UV continuum luminosity separate these two
spectra. Note, too, that the narrow emission line spectrum is
completely missing from the quasar composite. The blue wing of the Sy~1
Ly$\alpha$ profile is contaminated by geocoronal Ly$\alpha$, and in the
quasar composite spectrum the region shortward of the peak of
Ly$\alpha$ is affected by ``Ly$\alpha$ forest'' absorption. This is a
log-log plot.}
\end{figure}

Since Baldwin's initial discovery, several studies have found similar
relationships for UV emission lines in addition to C~IV, and that each
emission line has its own logarithmic slope in the $W_{\lambda} -
L_{uv}$ relation, perhaps offering clues to its origin (Baldwin,
Wampler, \& Gaskell 1989; Kinney, Rivolo, \& Koratkar 1990; Zamorani et
al.\ 1992; Espey, Lanzetta, \& Turnshek 1993; Osmer, Porter, \& Green
1994; Laor et al.\ 1995; Zheng, Kriss, \& Davidson 1995; Turnshek
1997). Kinney et al.\ (1990) also showed that the Baldwin effect
extends down to Seyfert nuclei luminosities.  Unfortunately, many of
these same studies also demonstrated the existence of substantial
intrinsic scatter in the relations. Thus, the use of quasars as
``beacons of cosmic history'' via the Baldwin effect would have to
await some understanding of quasar spectra.

\subsubsection{Other correlations.}

A number of other important quasar spectral correlations have been
amassed since Baldwin's discovery in 1977 and these may provide clues
to the latter's origin.

Quasars are intrinsically luminous X-ray sources. A number of quasar
X-ray emission studies have found a relationship between the UV/X-ray
continuum flux ratio \begin{equation} \frac{f_{uv}}{f_{x}} =
\left(\frac{\nu_{uv}}{\nu_{x}}\right)^{-\alpha_{uvx}} \end{equation}
and the UV continuum luminosity $L_{uv}$, in that higher luminosity
quasars have characteristically steeper effective logarithmic UV to
X-ray spectral slopes $\alpha_{ox}$ (2500~\AA\/ -- 2 keV) and
$\alpha_{uvx}$ (1350~\AA\/ -- 1 keV) (Zamorani et al.\ 1981; Wilkes et
al.\ 1994; Wang, Lu, \& Zhau 1998). The Wilkes et al.\ (1994) relation
is \begin{equation} \alpha_{ox} \approx 1.53 +
0.11\log\left(\frac{L_{\nu}(2500)}{10^{30}}\right). \end{equation} This
relation extends down to the luminosities of Seyfert nuclei. It should
be noted that the significance of this relationship is a subject of
debate (LaFranca, Franceschini, \& Vio 1995; Avni, Worrall, \& Morgan
1995; Yuan, Siebert, \& Brinkmann 1998).

Line ratios sensitive to the ionizing continuum SED, e.g.,
Ly$\alpha$/C~IV and Ly$\alpha$/O~VI, and the equivalent widths of these
three emission lines all correlate with $\alpha_{ox}$ and
$\alpha_{uvx}$. This is an expected result based upon energetics
arguments {\em if} $\alpha_{ox}$ and $\alpha_{uvx}$ are proxies for the
actual ionizing continuum SEDs of quasars (Netzer, Laor, \& Gondhalekar
1992; Zheng et al.\ 1995; Green 1996; Wang et al.\ 1998).  As explained
above these line ratios reflect the energy balance of the gas and thus
broadly measure the average ionizing photon energy of the incident
continuum. Relationships between $L_{uv}$ and these line ratios and the
line equivalent widths were also found in these studies.

Taken together, these correlations would indicate a general
relationship between the quasar's ionizing SED and its observed
ultraviolet luminosity, and this may play a major role in the origin of
the Baldwin effect.

\section{``Natural Selection'' and the emission line spectra of quasars}

\subsection{Emission line spectrum from a single cloud}

The emission lines of quasars arise mainly or exclusively via the
photoionization of gas, in the so-called broad line region, surrounding
a UV/X-ray ionizing continuum source.  Early successful photoionization
models of emission from single clouds are described in the works of
Davidson \& Netzer (1979) and Kwan \& Krolik (1981). The incident
continuum SED and gas chemical abundances are the basic ``raw
materials'' from which the emitted spectrum from a single cloud
arises.  For a given set of raw materials, the ionizing photon flux,
the gas density, and {\em especially} the ratio of these two parameters
--- proportional to the ionization parameter $U(H)$ --- set the
spectrum.  The gas column density and details concerning the radiation
transfer place the final touches on the emitted spectrum. The early
models found that there was a preferred value of $U(H) \approx 0.01$,
though a separate population of gas with a larger value must also exist
to account for the observed strength of the O~VI $\lambda$1034 line.

\subsection{An observational paradox and its resolution}

\begin{figure}
\plotfiddle{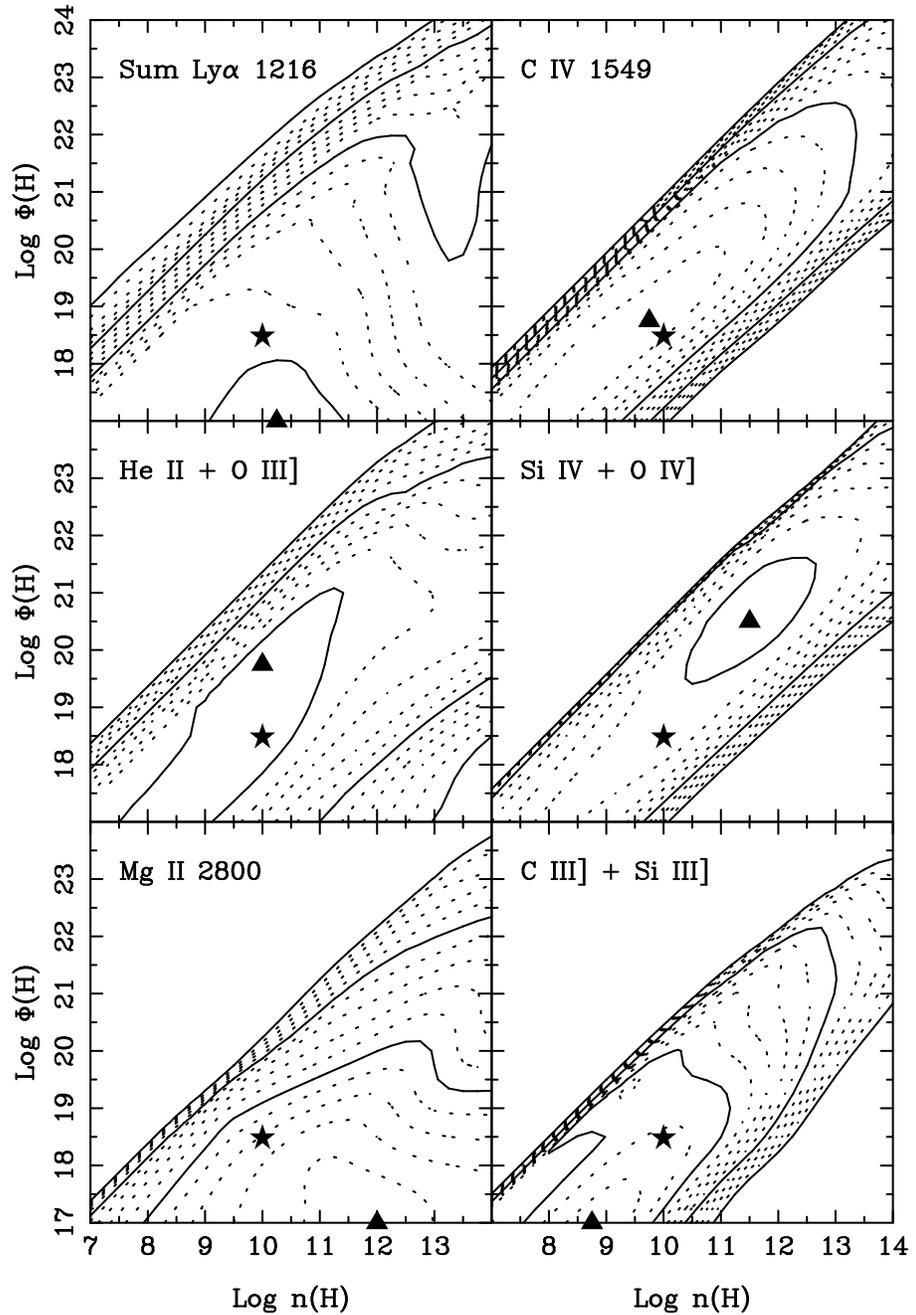}{6.0in}{0}{70}{70}{-205}{-30}
\caption{Contours of $\log W_{\lambda}$, referenced to the incident
continuum at 1215~\AA\/ and for full source coverage, for six emission
lines (or blends) are shown as a function of the hydrogen density and
flux of hydrogen ionizing photons. The total hydrogen column density is
$10^{23}~\rm{cm^{-2}}$. The $W_{\lambda}$ is in direct proportion to
the continuum reprocessing efficiency. The smallest (generally
outermost decade contour corresponds to 1~\AA\/, each solid line is 1
decade, and dotted lines represent 0.2 decade steps. The contours
generally decrease monotonically from the peak (solid triangle) to the
1~\AA\/ contour. The solid star is a reference point marking the old
``standard BLR'' parameters. ``Sum Ly$\alpha$~1216'' is the sum of
Ly$\alpha$~1216, He~II~1216, and O~V]~1218.}
\end{figure}

The work of many observational spectroscopists of the late 1970s and
early 1980s demonstrated the homogeneous nature of the emission line
spectra of quasars and their lower luminosity cousins, Seyfert nuclei.
Nature must somehow know to adjust the ionization parameter knob to the
nearly the same value through 5 decades in luminosity and 10 billion years
of lookback time.

From the mid 1980s to the mid 1990s, the monitoring of Seyfert nuclei
variability and emission line reverberation (Peterson 1993)
demonstrated that the broad line regions lie nearer to their continuum
sources than thought earlier, and contain gas with a range of particle
densities to accommodate a range in ionization through a region that
spans at least $\sim$~1.5 decades in radius. Rees, Netzer, \& Ferland
(1989) and Goad, O'Brien, \& Gondhalekar (1993) and others explored
models in which the gas density and column density were unique to and
allowed to vary with the distance from the ionizing photon source.
However, the means by which the quasar normalizes its gas properties or
why one power law of gas properties is chosen over another remained
unknown.  Somehow, over a huge range in ionizing continuum luminosity,
gas that is broadly distributed in space, particle density, and
ionization is able to produce the same sets of emission lines in
roughly similar ratios.  What's going on?

In the history of quasar photoionization models, the ``single cloud''
model was followed by the ``two cloud'' model, and then a radial power
law in cloud properties. Recently, Baldwin et al.\ (1995) extended the
``multi-cloud'' idea to its next logical step: ``Locally
Optimally-Emitting Clouds (LOC).'' What if nature provides the broad
line regions with a large pool of gas properties, such that at every
radius there exists a large range in particle and column densities?
Under these conditions the emitted spectrum is controlled by powerful
{\em selection effects} introduced by the atomic physics and basic
radiative transfer effects.  See Figure~2 and Korista, Baldwin, \&
Ferland (1997). Gas having properties that make it most efficient in
reprocessing the continuum into a particular line will emit most of
that line's luminosity. The gas need not fill the entire $n(H) -
\Phi\/(H)$ plane in Figure~2, but unless the gas in the BLR has a
remarkably restricted range of properties, we will in fact observe the
optimally emitting gas for most lines. And just as important, such an
ensemble spectrum reproduces a typical quasar spectrum.

The LOC model suggests that at work is a loose analog of Darwin's
natural selection and it offers natural explanations for: (1) the
observed homogeneity of the quasar spectra and the ``magic ionization
parameter'' problem, and (2) the ionization stratification inferred
from observations. Another important consequence of this idea is that
to a good approximation we can take a ``statistical mechanics''
approach to the modelling of quasar spectra, concerning ourselves with
the ensemble emission, rather than with the details of the properties
and arrangement of the individual gas clouds.  In this picture the
continuum SED and gas chemical abundances are the major drivers of the
emission line spectrum, not a magic ionization parameter.  In this case
we can hope to learn about these two fundamental parameters and the
possible stories they can tell concerning cosmic history and the nature
of quasars.

A word of caution should appear here. The emission line spectrum from
gas distributed in particle density and radius is in general much less
sensitive to changes in the gas abundances and continuum SED than that
emitted by a single ``cloud'' of fixed particle density and ionization
parameter.  This is because that as a cloud's thermodynamics adjusts to
changes in the SED or gas abundances, its reprocessing efficiency can
change immensely. In {\em any} distributed cloud model, there will be
other clouds with other properties at similar distances from the
ionizing source that will also react, and the integrated response is
dampened.  Under the distributed cloud hypothesis, the emitted spectrum
does remain sensitive to the continuum SED and gas abundances, and this
sensitivity will probably better represent nature.

\section{What are the parameters that govern the Baldwin Effect?}

I use the spectral synthesis code, Cloudy (90.04; Ferland et
al.\ 1998), and the emission from an ensemble of plane-parallel clouds
under the assumption of the LOC model to demonstrate that a
characteristic dependence of the continuum SED and gas abundances on
the quasar luminosity may provide an explanation for the Baldwin
effect. For simplicity and because of the spectrum's general
insensitivity to it, I held the column density fixed at $N(H) =
10^{23}~\rm{cm^{-2}}$. See Korista, Baldwin, \& Ferland (1998) for
details.

\subsection{The continuum SED}

For reasons of simplicity and because we do not know what ionizing
continuum SED is incident upon the broad line emitting gas, I will
demonstrate the effects of a changing continuum SED on the gas emission
using simple power laws, their spectral indeces $\alpha$ ranging from
$-2$ to $-1$ ($F_{\nu} \propto \nu^{\alpha}$).

Panel (b) of Figure~3 shows how the equivalent widths (here, always
measured with respect to the continuum at 1215~\AA\/) of 5 prominent
quasar emission lines vary with the power law index. The gas abundances
were fixed to roughly solar values in all cases. The emission line
equivalent widths ($W_{\lambda}$) all diminish with the softening in
the continuum. This is simply a result of conservation of energy ---
fewer high energy ionizing photons and the resulting lower gas
temperatures produce weaker lines relative to the {\em non-ionizing} UV
continuum at 1215~\AA\/. Panel (d) of Figure~3 shows that ratios
Ly$\alpha$/O~VI and Ly$\alpha$/C~IV are highly sensitive to the
continuum SED, but that N~V/He~II and especially N~V/C~IV are not.
Other SED-insensitive line ratios are $\lambda$1900~blend/C~IV,
$\lambda$1400~blend/C~IV, and N~III~990/C~III~977.

Figure~4 compares our simulations for fixed gas abundances to the
observations of Ly$\alpha$/O~VI (Zheng et al.\ 1995) and
Ly$\alpha$/C~IV (Wang et al.\ 1998). The observational points used
measurements of $\alpha_{ox}$ and $\alpha_{uvx}$, while of course the
simulations have assumed a simple power law with index $\alpha$. The
correspondence is good and demonstrates that the observed $\alpha_{ox}$
and $\alpha_{uvx}$ must stand as reasonably good proxies for the true
ionizing continuum SED.

Zheng \&  Malkan (1993) proposed that the Baldwin effect could be
explained via the combined effect of a strengthening non-ionizing UV
continuum and weakening ionizing continuum with luminosity.  That the
higher ionization lines in Figure~3b have steeper $W_{\lambda}$
dependencies on the spectral index is a reflection of these two
effects, and is generally consistent with the trends in the measured
slopes of the Baldwin effects of various emission lines (Espey et
al.\ 1993; Espey, these proceedings).

A SED--$L_{uv}$ relationship might be driving the Baldwin effect, but
it cannot be the whole story. When the $W_{\lambda}$(N~V) is measured
it has little or no dependence upon luminosity (Espey et al.\ 1993;
Osmer et al.\ 1994; Laor et al.\ 1995).  Figure~1 illustrates this
further. Panels (b) and (d) of Figure~3 predict that N~V/C~IV should be
luminosity independent, i.e., N~V should show a Baldwin effect as
strong as C~IV. Hamann \& Ferland (1993) showed that larger values of
N~V/C~IV and N~IV/He~II are systematically found at larger quasar
luminosities. This must mean that the continuum shape is not the only
parameter.

\begin{figure}
\plotfiddle{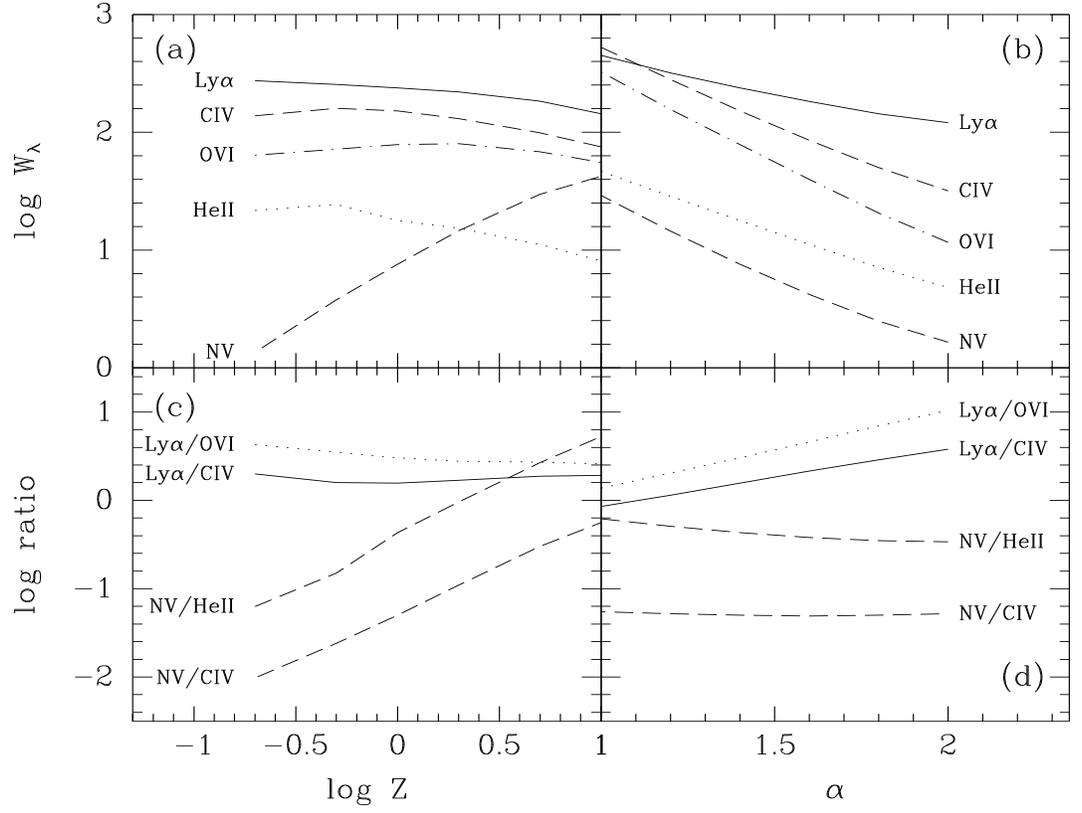}{4.0in}{270.}{80}{80}{-310}{410}
\caption{The equivalent widths of 5 prominent quasar emission lines and
their ratios as a function of metallicity ($Z$; panels [a], [c]) for a
fixed continuum SED ($\alpha = -1.4$), and as a function of the power
law index of the continuum SED ($\alpha$; panels [b], [d]) at a fixed
gas metallicity ($Z = 1$).}
\end{figure}

\begin{figure}
\plotfiddle{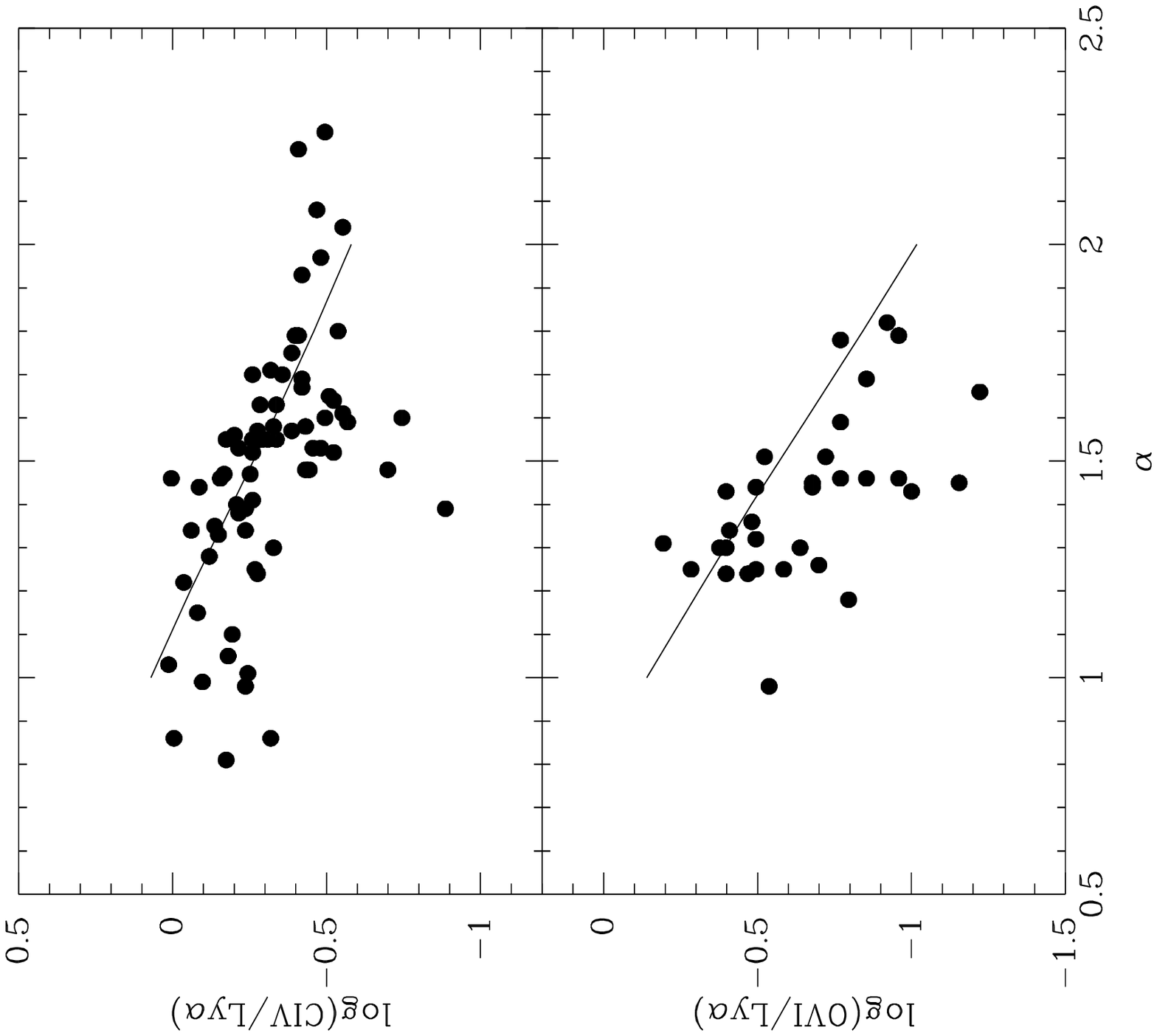}{4.0in}{270.}{80}{80}{-335}{450}
\caption{Observed emission line ratios vs.\ $\alpha_{ox}$ (solid dots)
and predicted emission line ratios vs.\ $\alpha$ (solid lines).}
\end{figure}

\subsection{Metallicity}

Here I will demonstrate the effects of a changing gas metallicity, $Z$,
for a fixed continuum SED ($\alpha = -1.4$). In a simple scheme to
change the metallicity I assumed that the metal abundances relative to
hydrogen scale from their solar values as $Z$. However, nitrogen is
expected to be selectively enhanced by ``secondary'' CNO processing in
stellar populations (at least when Z exceeds a few-tenths solar), and so
in these simulations $\rm{(N/H)} \propto Z^2$ while $\rm{(N/C)} \propto
Z$ and $\rm{(N/O)} \propto Z $ (Hamann \& Ferland 1993; Vila-Costas \&
Edmunds 1993; van Zee, Salzer, \& Haynes 1998). As such, nitrogen lines
are valuable tracers of the gas chemical enrichment (Ferland et
al.\ 1996).

In panel (a) of Figure~3 I show how the equivalent widths of the same
five emission lines vary with the metallicity. The strengths of
Ly$\alpha$ and especially He~II diminish with the increasing metal
opacity. In as far as hydrogen and helium are still the major opacity
sources, their ionized volumes diminish with increasing $Z$ due to the
increased numbers of electron donors. For $Z > Z_{\odot}$, the strengths
of C~IV and O~VI both diminish as the gas cools and the cooling shifts
to other lines.  However, since N~V is initially a weak line and
because the N/H abundance ratio is proportional to $Z^2$,
$W_{\lambda}$(N~V) $\propto Z$! Other initially weak lines are
generally correlated to a lesser extent with metallicity. However,
those lines that are formed near the front of the cloud (e.g., O~IV]
1402) remain roughly constant with $Z$. The temperatures here scale
strongly with $Z$ (i.e., cooler with increasing $Z$) and the He$^{++}$
zone where these lines form is diminshed in volume. In panel (c) of
Figure~3 I show the variations in the same five line ratios with
metallicity. For reasons given above, Ly$\alpha$/O~VI and
Ly$\alpha$/C~IV are relatively insensitive to the gas metallicity.
However, the N~V/He~II and N~V/C~IV ratios are roughly linear in $Z$.
We also find the following rough dependencies on metallicity:
$\lambda$1900/C~IV $\propto Z^{0.6}$, N~III~990/C~III~977 $\propto Z$,
and $\lambda$1400/C~IV $\propto Z^{0.3}$.

It is worth noting that the same line ratios that are insensitive to
the continuum SED are sensitive to $Z$, and those sensitive to the
continuum SED are insensitive to $Z$. While the relationship is not
always simple, this dichotomy of emission line ratio properties is not
a coincidence, and may prove to be a powerful tool in disentangling the
effects of these two fundamental parameters on the quasar spectrum.

\subsection{An explanation?}

Panels (a) and (b) of Figure~3 suggest a solution to the N~V
conundrum.  If, characteristically, the gas metallicity increases (with
N/H $\propto Z^2$) while the continuum SED softens with increasing
luminosity, the $W_{\lambda}$(N~V) will remain approximately constant
and N~V/C~IV and N~V/He~II will increase with luminosity, as observed.

The models do not tell us how to link the SED and $L_{uv}$ together, so
I adopted the empirical $\alpha_{ox} - L_{uv}$ relation found by Wilkes
et al.\ (1994; $H_o = 50~\rm{km~s^{-1}~Mpc^{-1}}$ and $q_o = 0$), and
adjusted slightly the logarithmic zero point to reference the continuum
at 1550~\AA\/. Guided by the N~V/He~II, N~V/C~IV, and $L_{uv}$
relations found by Hamann \& Ferland (1993) I guessed a $L_{uv} - Z$
relation: $Z \approx 1$ for the lowest luminosity quasars (i.e.,
Seyfert nuclei) and $Z \approx 10$ for the highest luminosity quasars.
In this way I adopted the following simple relation to represent the
hypothesis:  \begin{equation} \log Z =
0.183\log\left(\frac{L_{\nu}(1550)}{10^{30}}\right) + 0.477.
\end{equation} After coupling the adopted SED -- $L_{uv}$ and $Z$ --
$L_{uv}$ relations, the photoionization simulations then predict the
slopes, $\beta$, of the Baldwin effect for the different emission lines
and blends. We show these in Table~1, in descending order of the ions'
destruction ionization potentials.

\begin{table}
\caption{Simulated Baldwin relation slopes}
\begin{tabular}{lc}
\tableline
\multicolumn{1}{c}{Emission Line} & \multicolumn{1}{c}{$\beta$} \\
\tableline

O~VI $\lambda$1034 & -0.22 \\
N~V $\lambda$1240 & -0.047 \\
C~IV $\lambda$1549 & -0.20 \\
He~II $\lambda$1640 & -0.17 \\
$\lambda$1400 & -0.13 \\
$\lambda$1900 & -0.083 \\
Mg~II $\lambda$2800 & -0.092 \\
Ly$\alpha$~$\lambda$1216 & -0.10 \\
\tableline
\tableline
\end{tabular}
\end{table}

The predictions in Table~1 follow both the measured amplitudes and
trends of the Baldwin effect slopes reported in the literature
(Baldwin, Wampler, \& Gaskell 1989; Kinney, Rivolo, \& Koratkar 1990;
Zamorani et al.\ 1992; Espey, Lanzetta, \& Turnshek 1993; Osmer,
Porter, \& Green 1994; Laor et al.\ 1995; Zheng, Kriss, \& Davidson
1995; Turnshek 1997). Note, however, that it is difficult to compare
results from different works since the investigators assumed different
values of $q_o$ whose effect is non-uniform in the $W_{\lambda} -
L_{uv}$ plane. Variations in slope due to different assumptions of
$q_o$ are at the 0.04 decade level, comparable to the typically quoted
uncertainties in the measured slopes.

The interplay of the luminosity dependent SED -- $Z$ effects determines
the spectrum, and accounts for the differing Baldwin effect slopes
amongst the different emission lines. While it has been suggested that
the observed strengthening of the Ly$\alpha$/C~IV and
$\lambda$1900/C~IV ratios with luminosity implies a shift to lower
ionization parameters with increasing quasar luminosity (e.g.,
Mushotzky \& Ferland 1984), I would argue for characteristic shifts to
softer continuum SEDs and larger values of $Z$. The predicted changes
in Ly$\alpha$/C~IV with $L_{uv}$ are dominated by a shift in the
continuum SED, while changes in $\lambda$1900/C~IV are dominated by
changes in $Z$.

\section{Discussion}

While I have shown that a SED -- $Z$ -- $L_{uv}$ relationship holds
promise in explaining the origin of the Baldwin effect, the SED --
$L_{uv}$ and $Z$ -- $L_{uv}$ relations or at least their combined
effect must be firmly established and then calibrated, before we can
hope to understand what the Baldwin effect can tell us about cosmology
or cosmic history. I have presented several emission line ratios that
should provide the means to disentangle the effects of the continuum
SED and $Z$.

But why should we expect a SED -- $Z$ -- $L_{uv}$ relationship? We are
beginning to see quasars in the context of the environment of massive
galaxies ($\S~1$), rather than isolated, exotic beasts. The
co-evolution of massive galaxies and the quasars within them may offer
an answer to the above question, and thus may be the fundamental origin
of the Baldwin effect in quasar spectra. Below, I try to summarize this
emerging, yet, I caution, immature picture.

{\em 10 $>$ z $>$ 2.~} The rapid turn on of quasars was the result of
massive galaxy formation, during which the action of merging halos
(Press \& Schecter 1974) both created and drove gas into supermassive
black holes (Haehnelt, Natarajan, \& Rees 1998; Miller \& Percival
1998). Masses of the nuclear black holes, $M_{bh}$, correlated with
masses of forming galaxies or their dark halo masses (Haehnelt et
al.\ 1998).  More massive galaxies, with their deeper central
gravitational wells, were better able to retain gas deposited during
stellar evolution than were lower mass galaxies, so they had a more
rapid buildup of the heavy elements in their nuclear regions (Hamann \&
Ferland 1993; 1999).  The most massive black holes likely produced the
most luminous quasars, {\em initially}, since quasars turned on near or
above their Eddington limits:  $L/L_{Edd} \propto \dot{M}/M_{bh} \geq
1$. These most luminous quasars thus required the largest mass
accretion rates, $\dot{M}$, to sustain their luminosities.  While we
are far from understanding the mechanisms that generate the quasar
continua (see Antonucci 1988; 1998), the standard accretion models
predict softer UV bumps for larger $M_{bh}$ at a fixed $\dot{M}/M_{bh}
\propto L/L_{Edd}$ (Laor \& Netzer 1989; Sincell \& Krolik 1998), and
the current corona -- accretion disk models predict larger UV to X-ray
continuum flux ratios for larger values of $\dot{M}/M_{bh}$ (Czerny,
Witt, \& $\rm{\dot{Z}}$ycki 1996).  Therefore, more luminous quasars
might be expected to have typically softer continuum SEDs (e.g.,
steeper $\alpha_{ox}$).  With the largest required mass accretion
rates, the most massive and luminous quasars are expected to have the
shortest lifetimes (like massive stars), as the limited ``fuel'' supply
is exhausted.  The fuel exhaustion process is not well understood, but
there are some ideas. At sufficient luminosities relative to the disk
binding energy, a ``back reaction'' may occur, choking off the matter
accretion significantly (Haehnelt et al.\ 1998). Furthermore, the
self-similar infall solutions of Bertschinger (1985) and the
Press-Schecter theory of the collapse of dark matter halos (1974) both
predict a decline in the mass accretion rate with time. Even if
$\dot{M}$ were to remain constant, the ratio $\dot{M}/M_{bh}$ must
evolve to smaller values due to accretion alone. Thus the evolution of
an accreting black hole quite naturally leads to a drop in
$\dot{M}/M_{bh}$ with time (see also Haiman \& Menou 1998). Once
$\dot{M}/M_{bh}$ drops below a critical value, the accretion flow
radiation efficiency is expected to drop to values $\eta\/ \ll~0.1$,
and an advection dominated accretion flow (ADAF) may result (Narayan \&
Yi 1995).  In this mode, the UV to X-ray continuum luminosities are
expected to be more comparable. Thus, as $\dot{M}/M_{bh}$ in individual
quasars diminished with time, evolution to lower luminosities and
harder continuum SEDs (e.g., flatter $\alpha_{ox}$) likely resulted
(Caditz, Petrosian, \& Wandel 1991; Yi 1996; Haehnelt et al.\ 1998).

{\em z $<$ 2.~} By this time most of the massive galaxies had already
formed and so the major fire works were over --- most quasars had begun
their death march, dying of exhaustion. The peak in the {\em active}
black hole mass function likely slippled slowly to smaller values with
time (Siemiginowska \& Elvis 1997), and with it the corresponding
required $\dot{M}$, perhaps allowing for longer quasar lifetimes.  The
typical quasar SED would evolve as discussed above. The initial nuclear
gas metallicity would be characteristically smaller in the shallower
potential wells of the lower mass black holes. Further evolution to
smaller values in $Z$ might have occurred in quasars of any mass due to
dilution by lower $Z$ gas falling into the nuclear regions over the
life of the galaxy (Hamann \& Ferland 1993).  Later epoch galaxy
mergers could rekindle the nuclear fires and again enrich the gas,
temporarily extending the quasar epoch. But in a maturing and expanding
universe, galaxy merger rates slowed, and eventually all quasars died.
Remaining are the Seyfert nuclei with their typically smaller $M_{bh}$,
$\dot{M}$, and $\dot{M}/M_{bh}$, lying within the centers of more
modest galaxies, whose nuclear gas metallicities are within a factor of
2 of solar. Also present are the more massive, yet generally dormant,
black holes in the nuclei of present-day giant elliptical and other
galaxies (Margorrian et al.\ 1998).

The above picture is presently sketchy at best, and may be completely
wrong. Key uncertainties are: the lives of quasars and their energy
dissipation mechanism(s), galaxy formation and its relation to the
births of the quasars. If this picture has any validity it is easy to
see why the Baldwin effect has been such a hard nut to crack.
Different initial conditions and times of massive galaxy formation
resulted in separate quasar populations, characterized by $M_{bh}$,
$\dot{M}$, $\dot{M}/M_{bh}$, and $Z$, co-evolving in time. This
scenario suggests that these four parameters are fundamental to the
quasar in driving its continuum and emission line spectra.  Osmer,
Porter, \& Green (1994) demonstrated that the Baldwin effect is seen
globally across all redshifts {\em and} within narrow redshift
intervals. This would argue for the study of carefully selected samples
of quasars at many redshift or lookback time intervals.

Further sources of intrinsic scatter in any continuum -- emission line
correlations can arise from geometry -- orientation effects due to
possible anisotropies in the line and continuum emission (Netzer 1987;
Netzer et al.\ 1992), as well as short time scale continuum -- emission
line reverberation (Kinney et al.\ 1990; Pogge \& Peterson 1992;
Shields, Ferland, \& Peterson 1995). The latter should be less
important for the higher luminosity quasars.  These and other nuisances
should ``average out'' in large samples of objects --- and perhaps the
luminosity distances can be recovered through a process analogous to
``main sequence fitting'' of stellar clusters, as suggested by Jack
Baldwin at this meeting. For a thorough discussion of these issues, see
the conference proceedings contribution of Osmer \& Shields.

While the jury is still out as to whether the Baldwin effect in quasar
spectra can ever be used as a standard luminosity candle, it may be
telling us about the workings of the quasar central engine, the gas
enrichment in the nuclei of massive galaxies, and possibly about the
evolution of massive galaxies themselves.

\acknowledgments

Jack Baldwin and Gary Ferland are my collaborators in the work
presented here. I also acknowledge helpful discussions with Fred Hamann
and Joe Shields. I am grateful to the Baldwin family for their gracious
hospitality during my two week stay in La Serena.

\end{document}